# Stock Investment: The p-index Approach

Xinzhao Xie [a], Bopei Nie [a], Kuo-Ping Chang [a, *]

[a] Jinhe Center for Economic Research, Xi'an Jiaotong University, Xianning West Road, 28#, Xi'an, Shaanxi Province, 710049, China.

* Corresponding author.

*E-mail address*: kpchang@mx.nthu.edu.tw (K. P. Chang)

Stock Investment: The p-index Approach

## ABSTRACT

Risk level can be defined as the probability that a party will fail to meet a stated obligation or promise. This paper has used European put option to construct the p-index to measure an asset's risk level. The p-index measures the insurance fee for each insured dollar to guarantee that the asset achieves at least a $\delta$ rate of return on a specified future date. The empirical results of using the p-index have shown that for the fifty stocks of China's SSE 50 index during 2018-2023, with the fair price strategy, one-week and one-month holding periods can earn more, and among seven economic sectors, materials sector stocks generated highest annualized rates of return: 11.04% (one-week period), 11.93% (two-week period) and 10.18% (one-month period). With momentum and contrarian strategies of one-week holding period, the p-ratio-efficient-contrarian strategy produced the highest annualized rate of return (9.97%), followed by the p-index-inefficient-momentum strategy (9.01%) and the p-index-efficient-contrarian strategy (6.48%), the MCIRS method employing the p-index consistently delivered higher returns than its beta-based approach, and efficient (outperforming) stocks failed to sustain their momentum while inefficient (underperforming) stocks exhibited no mean reversion. It is also found that the p-index-efficient-contrarian strategy outperformed in low-sentiment (low-volume) regimes, while the p-index-inefficient-momentum strategy outperformed during high-sentiment (high-volume) periods. For the five hundred stocks of the US S&P 500 index during 2018-2023, it is found that efficient stocks sustained their momentum while inefficient stocks exhibited mean reversion. The p-index-efficient-momentum strategy produced the highest annualized rate of return (3.69%), followed by the p-ratio-inefficient-contrarian strategy (3.67%) and the beta-efficient-momentum strategy (3.48%).

## 1. Introduction

Risk level can be defined as the likelihood that a party will fail to meet a stated obligation or promise. In finance literature, risk is commonly quantified using metrics such as beta (derived from the Capital Asset Pricing Model, CAPM) or the variance of an asset's rate of return. Although the CAPM is theoretically appealing, its empirical performance has been unsatisfactory. Banz (1981) noted that the market risk premium is not the sole determinant of asset risk premiums, finding that small-cap stocks tend to yield higher returns than large-cap stocks. Basu (1983) found that the earnings-price ratio affects returns. Similarly, Pástor and Stambaugh (2003) found that expected returns are related to sensitivity to aggregate liquidity. Fama and French (1992) pointed out that, in addition to market risk, firm size and value factors also significantly influence stock returns. Fama and French (1993) further developed a three-factor model, incorporating market risk, the size factor, and the value factor. Their empirical results indicate that the CAPM fails to fully account for all sources of stock returns. Fama and French (1996) found that the tests do not support the most basic prediction of the CAPM: that average stock returns are positively related to market betas. Fama and French (2015) later proposed a widely used five-factor model incorporating market risk, size, value, profitability, and investment factors. DeBondt and Thaler (1985) demonstrated that stock prices exhibit significant long-term return reversals, while Jegadeesh and Titman (1993) showed that stock returns display substantial short-term momentum. Conrad and Kaul (1998) demonstrated that momentum and contrarian strategies have an equal likelihood of success, albeit over different horizons: momentum is usually profitable at the medium term, whereas contrarian strategies generate statistically significant profits in the long run. Cooper, Gutierrez, and Hameed (2004) found that the mean monthly momentum profit is positive following positive market returns and negative following negative market returns. They also found that up-market momentum reverses in the long run. Carhart (1997) finds that mutual fund performance shows little evidence of skillful persistence, attributes most performance patterns to exposure to market, size, value, and momentum factors, and identifies high expenses as a major cause of persistent underperformance. But these multi-factor models face two challenges: First, the economic interpretation of factors remains contentious (e.g., whether the value factor represents a risk premium or behavioral bias). Second, factor premiums generally exhibit time decay (Green et al., 2017), suggesting that factors constructed using traditional financial indicators may be prone to overfitting, casting doubt on their long-term robustness.

This paper has used European put option to construct the p-index to measure the underlying asset's risk level (Chang, 2021 and 2023). Every asset which provides an uncertain payoff has a corresponding put-call parity. The p-index measures the insurance fee for each insured dollar to guarantee that the asset achieves at least a $\delta$ rate of return on a specified future date. Applying this p-index framework, the study evaluates the performance of various investment strategies across the fifty constituent stocks of China's SSE 50 index and the five hundred stocks of the US S&P 500 index over the period 2018-2023. The empirical results of China's fifty stocks show that with the fair price strategy, one-week and one-month holding periods can earn more than two-week and two-month holding periods, and among seven economic sectors, materials sector stocks generated highest annualized rates of return: 11.04% (one-week period), 11.93% (two-week) and 10.18% (one-month period). With (one-week holding period) momentum and contrarian strategies, the p-ratio-efficient-contrarian strategy produced the highest annualized rate of return (9.97%), followed by the p-index-inefficient-momentum strategy (9.01%) and the p-index-efficient-contrarian strategy (6.48%); the minimum convex input requirement set (MCIRS) method employing the p-index consistently delivered higher returns than its beta-based approach; and efficient (outperforming) stocks failed to sustain their momentum, while inefficient (underperforming) stocks exhibited no mean reversion. For the five hundred stocks of the US S&P 500 index, it is found that efficient stocks sustained their momentum while inefficient stocks exhibited mean reversion.

The remainder of this paper is organized as follows. Section 2 derives the p-index and its upper and lower bounds. It also discusses the properties of the p-index under the binomial asset pricing model. Data and empirical results are presented in Section 3. Concluding remarks appear in Section 4.

## 2. The p-index for Measuring Risk Structures of Assets

Assume that an asset's current price at $t = 0$ is $S_0$, and its value at $t = T$ is an uncertain $S_T$. For this asset, at $t = 0$ buying a European call option $c$ with the strike price $K$ will give the owner of the call: $Max[S_T - K, 0]$ at $t = T$. At $t = 0$ buying a European put option $p$ with the strike price $K$ will give the owner of the put: $Max[K - S_T, 0]$ at $t = T$. Thus, both $p$ and $c$ can be interpreted as insurance. The put option $p$ can be interpreted as: If at $t = 0$ a person buys both the underlying asset $S_0$ and an insurance $p$, then at $t = T$ the value of her owning the underlying asset will be worth at least $K$, i.e., $Max[S_T, K]$. That is, by paying $p$ as an insurance fee to an insurance company at $t = 0$, the insurance company will ensure that the underlying asset can deliver at least $K$ at $t = T$: $Max[S_T, K]$. Higher $p$ means less likelihood that the underlying asset will be worth at least $K$. The call option $c$, on the other hand, can be interpreted as: If at $t = 0$ a person short-sells the underlying asset $S_0$ and buys an insurance $c$ (where $K$ can be interpreted as the insurance deductible), then at $t = T$ this person will not need to pay more than $K$ to buy back the asset. That is, at $t = T$, to buy back the underlying asset this person will pay $Min[S_T, K]$ and the insurance company will pay $Max[S_T - K, 0]$ such that $S_T = Min[S_T, K] + Max[S_T - K, 0]$.

For any asset which delivers an uncertain payoff $S_T$ at $t = T$, there exists a corresponding put-call parity at $t = 0$:[1]

$$c + \frac{K}{1+r} = S_0 + p. \tag{1}$$

In eq. (1), if $K > (<)(1 + r)S_0$, then $p > (<)c$. If $K = (1 + r)S_0$, then $p = c$. That is, for the insurance company, when $K = S_0(1 + r)$, the risk of insuring that the underlying asset is worth at least $K$ at $t = T$ is equivalent to the risk of insuring that the insurant needs not to pay more than $K$ to buy the asset at $t = T$.[2]

Suppose that there are $n$ uncertain assets at $t = 0$. We can rewrite eq. (1) as:

$$S_{0i} = c_i + \frac{K_i}{1+r} - p_i, \quad i = 1, \dots, n,$$

and divide both sides of the above equation by $S_{0i}$:

---

[1] Consider two portfolios at $t = 0$ and a simple risk-free interest rate $r$:

Portfolio A: one European call option $c$ with strike price $K$, and cash $\frac{K}{1+r}$ deposited in a bank;

Portfolio B: one European put option $p$ with strike price $K$, and one unit of the underlying asset $S_0$.

On the expiration date $t = T$, both portfolios give exactly the same payoff: $Max[S_T, K]$. Thus, the costs of the two portfolios at $t = 0$ must be the same, i.e., $c + \frac{K}{1+r} = S_0 + p$.

[2] We can rewrite eq. (1) as: $S_0 = c + \left(\frac{K}{1+r} - p\right)$. If $S_0$ is the market value of a levered firm, then $c$ is the equity, $\left(\frac{K}{1+r} - p\right)$ is the risky debt, and $p$ is the insurance to ensure the promised payment $K$ to debtholders at $t = T$. Note that the changes of $K$ will not affect $S_0$, i.e., the Modigliani-Miller capital structure irrelevancy proposition is an example of 'financial diversification irrelevancy', see Chang (2023). With $K > 0$ and $S_0 > 0$, the risky debt must be strictly positive, i.e., $\left(\frac{K}{1+r} - p\right) > 0$, or $\frac{K}{1+r} > p$ and $\frac{K}{1+r} \neq p$.

$$1 = \frac{c_i}{S_{0i}} + \frac{K_i/S_{0i}}{1+r} - \frac{p_i}{S_{0i}}, \quad i = 1, \dots, n. \tag{2}$$

Let $K_i = S_{0i}(1+\delta)$ where $\delta > -1$, we have:

$$1 = \frac{c_i}{S_{0i}} + \frac{1+\delta}{1+r} - \frac{p_i}{S_{0i}}, \quad i = 1, \dots, n. \tag{3}$$

That is, for the i-th one-dollar asset (e.g., i-th one-dollar stock) at $t = 0$, to ensure that at $t = T$, this asset will deliver at least a $\delta$ rate of return (i.e., one-dollar becomes at least $1+\delta$ dollars), the insurance fee paid at $t = 0$ is: $\frac{p_i}{S_{0i}}$. Hence, if $\frac{p_i}{S_{0i}} > \frac{p_j}{S_{0j}}$, it means that the risk level of the i-th asset delivering at least a $\delta$ rate of return is greater than that of the j-th asset. We can rearrange eq. (3) as:

$$\frac{1}{1+\delta} = \frac{c_i}{S_{0i}(1+\delta)} + \frac{1}{1+r} - \frac{p_i}{S_{0i}(1+\delta)}, \quad i = 1, \dots, n, \tag{4}$$

and define the p-index as:

$$\text{p-index:} \quad \frac{p_i}{K_i} = \frac{p_i}{S_{0i}(1+\delta)}, \quad i = 1, \dots, n. \tag{5}$$

The p-index measures the insurance fee for each insured dollar to guarantee that the asset achieves at least a $\delta$ rate of return at $t = T$. Thus, higher p-index means higher risk (i.e., less likelihood) for the asset $S_0$ to deliver at least a $\delta$ rate of return at $t = T$.

As shown by Chang (2020, 2023), upper and lower bounds for the put option $p$ is: $Max\left[\frac{K}{1+r} - S_0, 0\right] \le p < \frac{K}{1+r}$, i.e., an asset cannot sell for more than or equal to the present value of a sure payment of its maximum payoff. Hence, upper and lower bounds for the p-index is: $Max\left[\frac{1}{1+r} - \frac{1}{1+\delta}, 0\right] \le \frac{p}{S_0(1+\delta)} < \frac{1}{1+r}$ where $K = S_0(1+\delta)$. Also, $\frac{\partial}{\partial K}\left(\frac{p}{K}\right) > 0$ if $\frac{\partial p/p}{\partial K/K} > 1$.[3]

**The Binomial Case**

[3] See also Chang (2021).

As shown in Figure 1, at $t = 0$, we have a stock with current price $S_0$ and a European call $c$ with exercise price $K$ and a European put option $p$ with exercise price $K$. Suppose that at $t = T$, the stock price will either go up to $S_0 u$ or go down to $S_0 d$, and $r$ is the simple risk-free rate of interest.

$$c_u = Max[S_0 \cdot u - K, 0]$$
$$p_u = Max[K - S_0 \cdot u, 0]$$

$$S_0 \rightarrow S_0 \cdot u$$
$$S_0 \rightarrow S_0 \cdot d$$

$c = ?$ $\quad c_d = Max[S_0 \cdot d - K, 0]$

$p = ?$ $\quad p_d = Max[K - S_0 \cdot d, 0]$

Figure 1. A binomial option pricing model

Then from the Arbitrage (Gordan) theory we have:[4]

[4] As shown in Chang (2015), the Arbitrage (Gordan) theory is:

Let $\boldsymbol{A}$ be an $m \times n$ matrix. Then, exactly one of the following systems has a solution:

System 1: $\boldsymbol{Ax} > \boldsymbol{0}$ for some $\boldsymbol{x} \in \boldsymbol{R^n}$

System 2: $\boldsymbol{A^T p} = \boldsymbol{0}$ for some $\boldsymbol{p} \in \boldsymbol{R^m}$, $\boldsymbol{p} \geq \boldsymbol{0}$, $\boldsymbol{e^T p} = 1$ where $\boldsymbol{e} = \begin{bmatrix} 1 \\ 1 \\ \cdot \\ \cdot \\ \cdot \\ 1 \end{bmatrix}$.

If System 2 holds (i.e., no arbitrage) and the matrix $\boldsymbol{A}$ has rank $m - 1$ (i.e., a complete market), the probability measure $\boldsymbol{p}$ will be unique. (Suppose that $\boldsymbol{A^T p} = \boldsymbol{0}$ and $\boldsymbol{p} = (\pi_1, \ldots, \pi_m)^{\boldsymbol{T}}$ is a non-zero vector. Then the rank of $\boldsymbol{A^T}$, $R(\boldsymbol{A^T})$, is less than $m$. Unique solution for $(\pi_1, \ldots, \pi_m)$ and $\sum_{i=1}^{m} \pi_i = 1$ imply $R(\boldsymbol{A^T}) = m - 1$). Note that System 2 is a martingale. Cox et al.'s (1979) binomial option pricing model is System 2 when System 1 does not hold, i.e., when there is no arbitrage.

$$\begin{bmatrix} (1+r)1-(1+r)1 & (1+r)1-(1+r)1 \\ S_0u-(1+r)S_0 & S_0d-(1+r)S_0 \\ Max(S_0u-K,0)-(1+r)c & Max(S_0d-K,0)-(1+r)c \\ Max(K-S_0u,0)-(1+r)p & Max(K-S_0d,0)-(1+r)p \end{bmatrix} \begin{bmatrix} \pi \\ 1-\pi \end{bmatrix} = \begin{bmatrix} 0 \\ 0 \\ 0 \\ 0 \end{bmatrix},$$

$$\boldsymbol{A^T} \qquad\qquad \boldsymbol{p} \quad = \boldsymbol{0}$$

or

$$\begin{cases} \text{Money market:} \quad 1 = \frac{1}{1+r}[\pi(1+r)+(1-\pi)(1+r)] \\ \text{The stock:} \quad S_0 = \frac{1}{1+r}[\pi \cdot S_0u + (1-\pi)\cdot S_0d] \\ \text{Call option:} \quad c = \frac{1}{1+r}[\pi \cdot Max(S_0u-K,0) + (1-\pi)\cdot Max\ (S_0d-K,0)] \\ \text{Put option:} \quad p = \frac{1}{1+r}[\pi \cdot Max(K-S_0u,0) + (1-\pi)\cdot Max\ (K-S_0d,0)] \end{cases} \tag{6}$$

where $\pi = \frac{(1+r)-d}{u-d}$ , $1-\pi = \frac{u-(1+r)}{u-d}$ , and $\boldsymbol{p} = (\pi, 1-\pi)$ is the probability measure governing the stochastic process. Thus, let $K = S_0(1+\delta)$, the p-index for the asset $S_0$ is:

$$\frac{p}{K} = \frac{\frac{1-\pi}{1+r}[S_0(1+\delta)-S_0d]}{(1+\delta)S_0} = \frac{(1+\delta)-d}{1+\delta}\cdot\frac{1-\pi}{1+r} \ . \tag{7}$$

It shows that if at $t=0$ an asset $S_0$ has a lower down move $d$, it implies higher risk (i.e., a lower likelihood) for the asset to achieve at least a $\delta$ rate of return at $t=T$, i.e., $\frac{\partial}{\partial d}\left[\frac{p}{(1+\delta)S_0}\right] = \frac{-1}{1+\delta}\cdot\frac{1-\pi}{1+r} < 0$, $\frac{\partial^2}{\partial d^2}\left[\frac{p}{(1+\delta)S_0}\right] = 0$; and $\frac{\partial}{\partial \delta}\left[\frac{p}{(1+\delta)S_0}\right] = \frac{d}{(1+\delta)^2}\cdot\frac{1-\pi}{1+r} > 0$, $\frac{\partial^2}{\partial \delta^2}\left[\frac{p}{(1+\delta)S_0}\right] < 0$. Also, assets having $d=0$ will have exactly the same constant p-index: $\frac{p}{K} = \frac{p}{(1+\delta)S_0} = \frac{1-\pi}{1+r}$ .

## 3. Data and Empirical Results

Daily fifty stocks data of the SSE 50 Index of 2018–2023 were adopted from the China Stock Market & Accounting Research Database (CSMAR Data). The SSE 50 Index selects 50 of the most representative stocks with large market capitalization and high liquidity from the Shanghai Securities Market to form its component stocks, aiming to comprehensively reflect the overall performance of leading enterprises with the most significant market influence in the Shanghai Securities Market. The SSE 50 Index was officially launched on January 2, 2004. Its objective is to establish an actively traded, large-scale investment index that primarily serves as the underlying basis for derivative financial instruments.

In a certain week, as in eq. (6), we can have:

$$\begin{cases} \text{Shanghai Composite Index (SSEC):} \ \widehat{S_{0SH}} = \frac{1}{1+r}[\pi \cdot S_{0SH}u_{SH} + (1-\pi) \cdot S_{0SH}d_{SH}] \\ \text{The i}-\text{th stock:} \ S_{0i} = \frac{1}{1+r}[\pi \cdot S_{0i}u_i + (1-\pi) \cdot S_{0i}d_i] \\ \text{Put of the i}-\text{th stock:} \ p_i = \frac{1}{1+r}[\pi \cdot Max(K_i - S_{0i}u_i, 0) + (1-\pi) \cdot Max(K_i - S_{0i}d_i, 0)] \end{cases} \tag{8}$$

where $i = 1, \ldots, 50$, and

1. $\widehat{S_{0SH}}$ is the closing price on Monday of Shanghai Composite Index (SSEC), $S_{0SH}u_{SH}$ is the highest closing price of SSEC of the week, $S_{0SH}d_{SH}$ is the lowest closing price of SSEC of the week, and $r$ is equal to the coupon rate on the 10-year China government bond divided by 52. Thus, we can calculate $\pi$ and $1-\pi$. If it is found that $1 > \pi > 0$ does not hold, then use this week's and the previous week's closing prices to calculate $\pi$.
2. For the i-th company ($i = 1, \ldots, 50$), $S_{0i}u_i$ is the highest closing price of the i-th stock of the week, and $S_{0i}d_i$ is the lowest closing price of the i-th stock of the week. With $r$, $\pi$, $S_{0i}u_i$ and $S_{0i}d_i$, we can estimate $S_{0i}$, the i-th stock's theoretical (fair) closing price on Monday of the week.
3. Let $K_i = (1+\delta)S_{0i}$ and $\delta = r$. With $K_i$, $r$, $\pi$, $S_{0i}u_i$ and $S_{0i}d_i$, we can estimate $p_i$, the i-th stock's put option closing price on Monday of the week. The p-index of the i-th stock in this week is: $\frac{p_i}{K_i} = \frac{p_i}{(1+r)S_{0i}}$ , which is the insurance fee for each insured dollar so that the i-th stock can achieve at least $r$ rate of return at the end of the week.

After deriving the p-index (i.e., risk level) of an asset over a five-day period (Monday through Friday), we can also calculate the p-ratio of the asset (with $\delta = r$) over that period:

$$\text{p-ratio:} \quad \frac{r_i - r}{\text{p-ind}} = \frac{r_i - r}{p_i/[S_{0i}(1+r)]}, \quad i = 1, \ldots, 50, \tag{9}$$

where $r_i$ is the rate of return of the i-th stock in the week. The p-ratio, as the Treynor ratio, can be defined as an asset's risk-adjusted excess return, i.e., how much return an asset investment earned for the amount of risk the asset investment assumed.

Figure 1 shows an example of SAIC (Shanghai Automotive Industry Corporation) Motor's weekly p-indexes (red line) and p-ratios (blue line) during 2018-2023. It shows that during 2020-2021 COVID-19 pandemic period the firm's p-index (risk-level) largely increased though the p-ratio (risk-adjusted excess rate of return) did not change significantly.

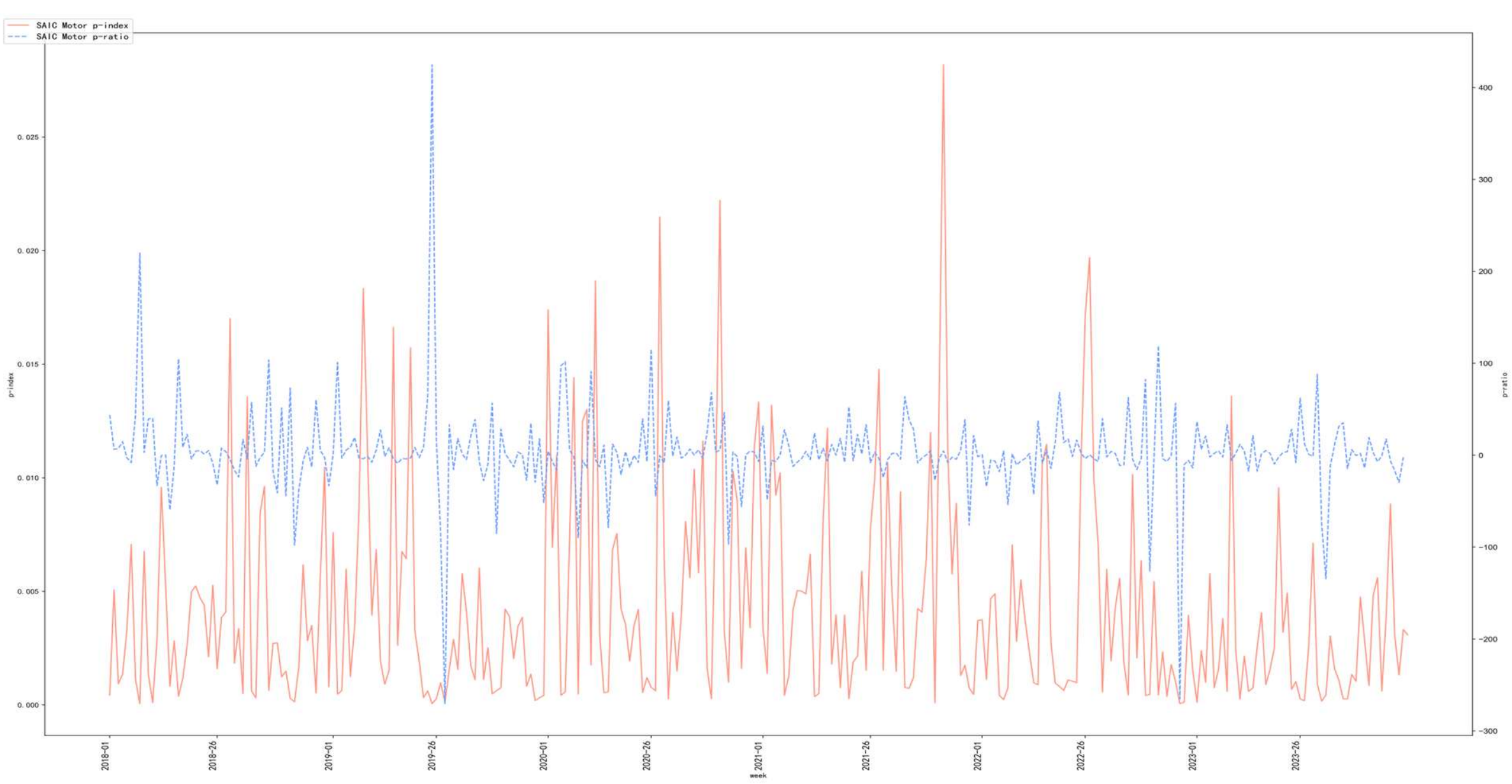

Figure 1 SAIC Motor's p-indexes and p-ratios during 2018-2023

### 3.1 Investment with the Fair Price Strategy

From eq. (8) we can calculate the i-th firm's theoretical (fair) stock price:

$$S_{0i} = \frac{1}{1+r}[\pi \cdot S_{0i}u_i + (1-\pi) \cdot S_{0i}d_i],\ i = 1, \dots, 50.$$

In weekly investments, a 3-week cycle serves as an example. Using the five closing prices from Week 1 (Monday through Friday), we calculate the fair Monday closing price, say, $S_{0i} = \$12$.

(i) Suppose the actual Monday closing prices over the three weeks are: $\widehat{S_{0\iota}} = \$9$ (Week 1), $\widehat{S_{0\iota}} = \$10$ (Week 2), and $\widehat{S_{0\iota}} = \$11$ (Week 3). Since in Week 1, $S_{0i} > \widehat{S_{0\iota}}$ (i.e., $\$12 > \$9$), the stock is undervalued. Thus, on the Monday of Week 2, we invest $\$100$ at the closing price of $\$10$, purchasing $10$ shares $(= 100/10)$. On the Monday of Week 3, we sell the 10 shares at the closing price of $\$11$, receiving $110. The rate of return for this weekly investment is: $10\% = \frac{110-100}{100}$ .

(ii) Suppose the actual Monday closing prices over the three weeks are: $\widehat{S_{0\iota}} = \$14$ (Week 1), $\widehat{S_{0\iota}} = \$10$ (Week 2), and $\widehat{S_{0\iota}} = \$15$ (Week 3). Since in Week 1, $S_{0i} < \widehat{S_{0\iota}}$ (i.e., $\$12 < \$14$), the stock is overvalued. Thus, on the Monday of Week 2, we short-sell $10$ shares at the closing price of $10, receiving $100. On the Monday of Week 3, we buy back $10$ shares at the closing price of $\$15$, spending $\$150$. The rate of return for this weekly investment is: $-50\% = (100 - 150)/100$.

For each stock, cumulative and annualized rates of returns (2018–2023) are derived by aggregating weekly rates of return. Annualized rates of return can be further calculated for two-week, one-month, and two-month holding periods. Annualized rates of return of equally weighted portfolios of SSE 50, three largest weights stocks, and materials sector stocks are shown in Table 1.

(1) For the equally weighted portfolio of the fifty stocks in the SSE 50 index, one-week and one-month holding periods have performed better than two-week and two-month holding periods.

(2) For the top three stocks with the highest weightings account for about 33% of the SSE 50 Index (one liquor company and two financial companies), one-week and one-month holding periods have performed well.

(3) Among the seven sectors in the SSE 50 Index (financial, healthcare, industrial, consumer, materials, technology, and energy), the materials sector stocks have outperformed the others. For the equally-weighted portfolio comprising seven companies focused on raw materials, chemicals, and green technology, all holding periods—with the exception of the two-month period—have delivered annualized returns exceeding 10%.

Table 1 Annualized rates of return of equally-weighted portfolios under different investment periods

| | Weekly | Two weeks | Monthly | Two months |
|---|---|---|---|---|
| SSE 50 stocks | 0.91% | 0.71% | 4.92% | -3.03% |
| Three largest weights | 2.31% | -10.24% | 9.78% | -1.25% |
| Materials sector | 11.04% | 11.93% | 10.18% | -1.02% |

We also employ the median test and Wilcoxon test to analyze differences in the p-index (the risk level) between high-yield companies (top 10% in cumulative returns) and low-yield companies (bottom 10% in cumulative returns) across four holding periods: one week, two weeks, one month, and two months.[5] It is found that in these four holding periods, the p-index exhibits statistically significant differences between high- and low-yield firms, and high-yield firms have lower p-index, and low-yield firms have higher p-index (Table 2).

Table 2 p-values for the differences in the p-index between high- and low-yield companies

| | Weekly | Two weeks | Monthly | Two months |
|---|---|---|---|---|
| Median test | 0.014088 | 0.543472 | 0.004586 | 0.002811 |
| Wilcoxon test | 0.001684 | 0.016622 | 0.000006 | 0.000057 |

### 3.2 Investments with Momentum and Contrarian Strategies

**Using p-ratios**

We can use the weekly p-index of each stock to calculate the weekly p-ratio of each stock, i.e., eq. (9). For a 3-week cycle,

(i) In Week 1, classify the top 10% of firms by p-ratio as efficient firms and the bottom 10% as inefficient firms.

(ii) For an efficient firm's stock, two strategies are applied:

a. Momentum strategy: Buy at the Monday closing price of Week 2 and sell at the Monday closing price of Week 3.

b. Contrarian strategy: Short sell at the Monday closing price of Week 2 and buy back (i.e., cover) at

[5] Since the p-index has upper and lower bounds, i.e., $Max\left[\frac{1}{1+r}-\frac{1}{1+\delta},0\right] \leq \frac{p}{S_0(1+\delta)} < \frac{1}{1+r}$ , non-parametric median test and Wilcoxon test are more appropriate.

the Monday closing price of Week 3.

For an inefficient firm's stock, two strategies are applied:

c. Contrarian strategy: Buy at the Monday closing price of Week 2 and sell at the Monday closing price of Week 3.

d. Momentum strategy: Short sell at the Monday closing price of Week 2 and buy back at the Monday closing price of Week 3.

As illustrated in Figure 2, the equally weighted p-ratio-inefficient portfolio using a momentum strategy did not perform well, generating annualized rate of return of just $1.17\%$. The equally weighted p-ratio-inefficient portfolio using a contrarian strategy is worse, providing annualized rate of return of $-1.24\%$.[6] In contrast, for equally weighted p-ratio-efficient portfolios, the contrarian strategy (Figure 3) delivered strong performance ($9.97\%$ annualized rate of return), while the momentum strategy incurred substantial losses ($-21.62\%$ annualized rate of return). These results suggest that in China's A-share market, efficient (outperforming) stocks fail to sustain their momentum, and inefficient (underperforming) stocks do not revert to an efficient state.

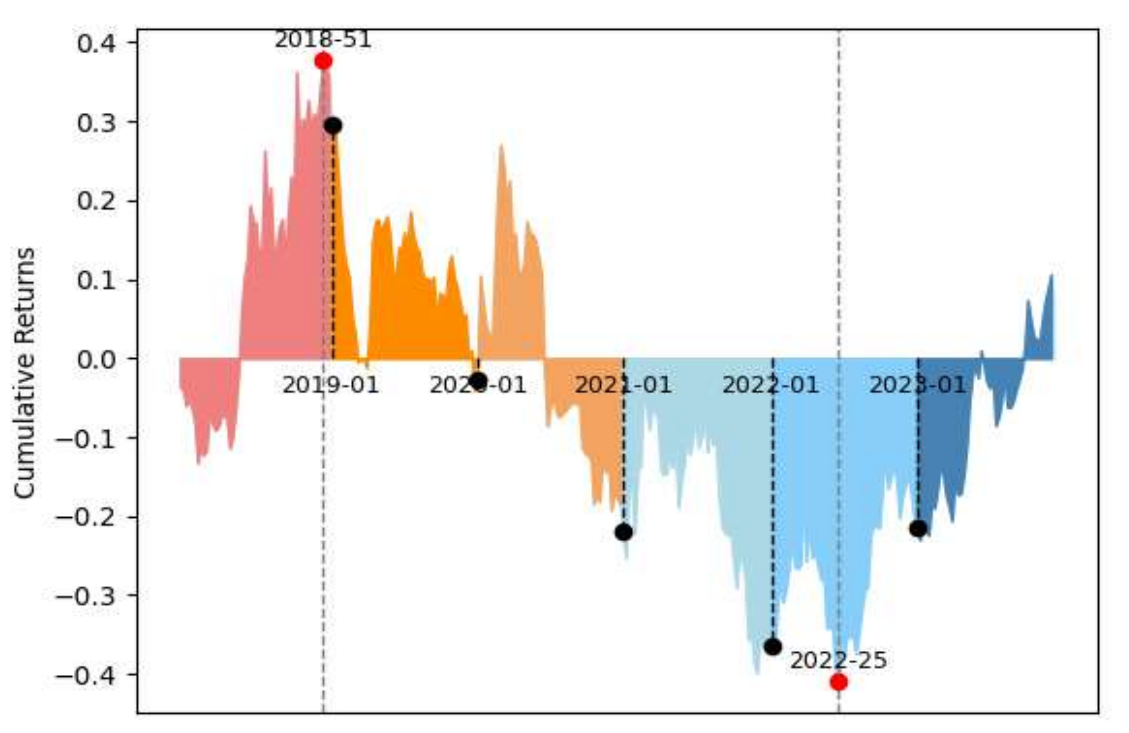

Figure 2 SSE 50 p-ratio-inefficient-momentum

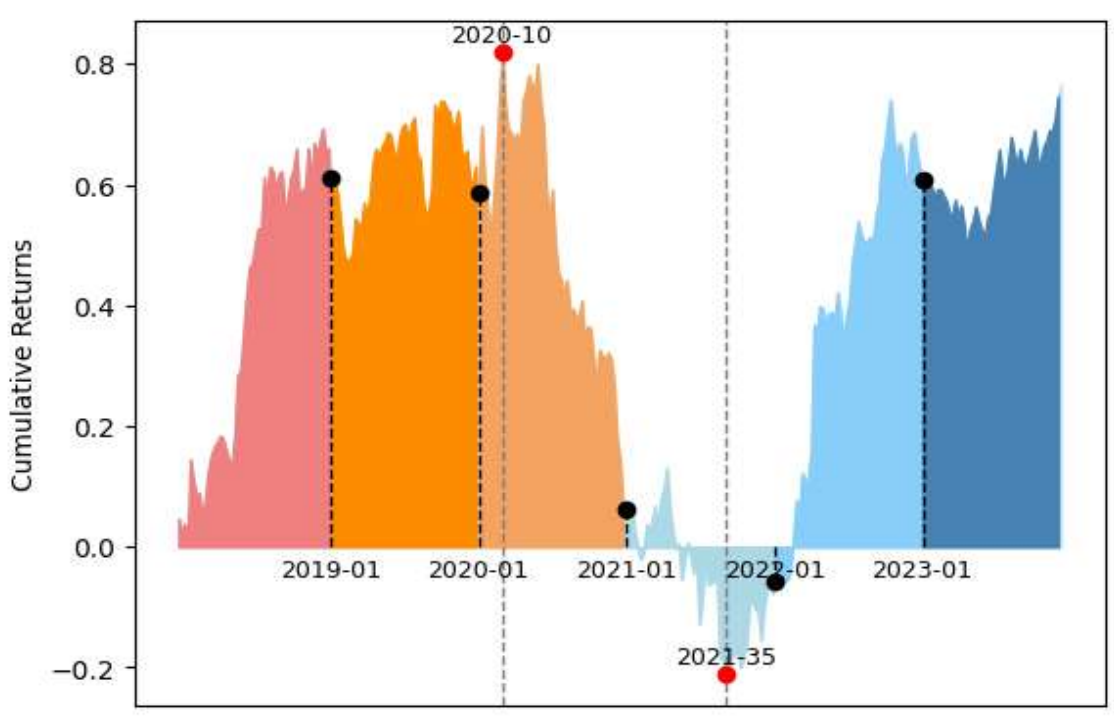

Figure 3 SSE 50 p-ratio-efficient-contrarian

## Using the MCIRS Approach

6 Note that the graph of momentum strategy is the vertically inverted version of the contrarian strategy.

Schwert (1983) found that factors affecting stock performance extend far beyond risk. Elements such as company size, capital stock, and number of employees also exert a significant influence on performance. Chang (1999, 2004) have developed the minimum convex input requirement set (MCIRS) approach. The MCIRS approach constructs a quasi-concave function to evaluate whether a decision-making-unit (e.g., a firm) uses inputs (e.g., beta, capital and labor) efficiently to produce output (e.g., rate of return). The approach is non-parametric, i.e., it does not employ any particular function to measure efficiency (i.e., no assumptions are made concerning the underlying production technology). The following linear programming is used to evaluate the k-th stock's efficiency in each week:

$$\begin{aligned} &\text{Maximize} \quad u_k \\ &\text{subject to} \quad \textstyle\sum_{i=1}^{3} v_i x_{ik} = 1 \\ &\qquad \textstyle\sum_{i=1}^{3} v_i x_{ij} - u_k \geq 0 \\ &\qquad \text{for all } j \in \overline{M_k} = \{j' : 1 \leq j' \leq 50;\ y_{j'} \geq y_k\} \\ &\qquad u_k, v_i \geq 0, \quad i = 1, 2, 3, \end{aligned} \tag{10}$$

where $u_k$ is the efficiency score ($0 \leq u_k \leq 1$), and one is the highest score, and zero, the lowest. $u_k < 1$ means that the k-th inefficient stock could be efficient if its three inputs, $x_{ik}$, $i = 1, 2, 3$, were reduced proportionally to $u_k x_{ik}$, $i = 1, 2, 3$, to produce the same output level $y_k$. In eq. (10), data of the three inputs (i.e., capital, labor, and beta) and the output (i.e., rate of return) were adopted from Choice Data (East Money Information).[7] In each week, after calculating the efficiency score for each stock, the top 10% of firms ranked by their efficiency scores are classified as efficient firms, while the bottom 10% are designated as inefficient firms.

Figure 4 shows that the equally weighted beta-inefficient portfolio using a momentum strategy performed well, generating annualized rate of return of $7.21\%$. The equally weighted beta-inefficient portfolio using a contrarian strategy incurred losses, providing annualized rate of return of $-11.47\%$. For equally weighted beta-efficient portfolios, the contrarian strategy (Figure 5) delivered strong performance ($5.98\%$ annualized rate of return), while the momentum strategy incurred substantial losses ($-8.6\%$ annualized rate of return).

[7] Beta, a measure of systematic risk relative to the market, is calculated using stock price data over a 100-week period.

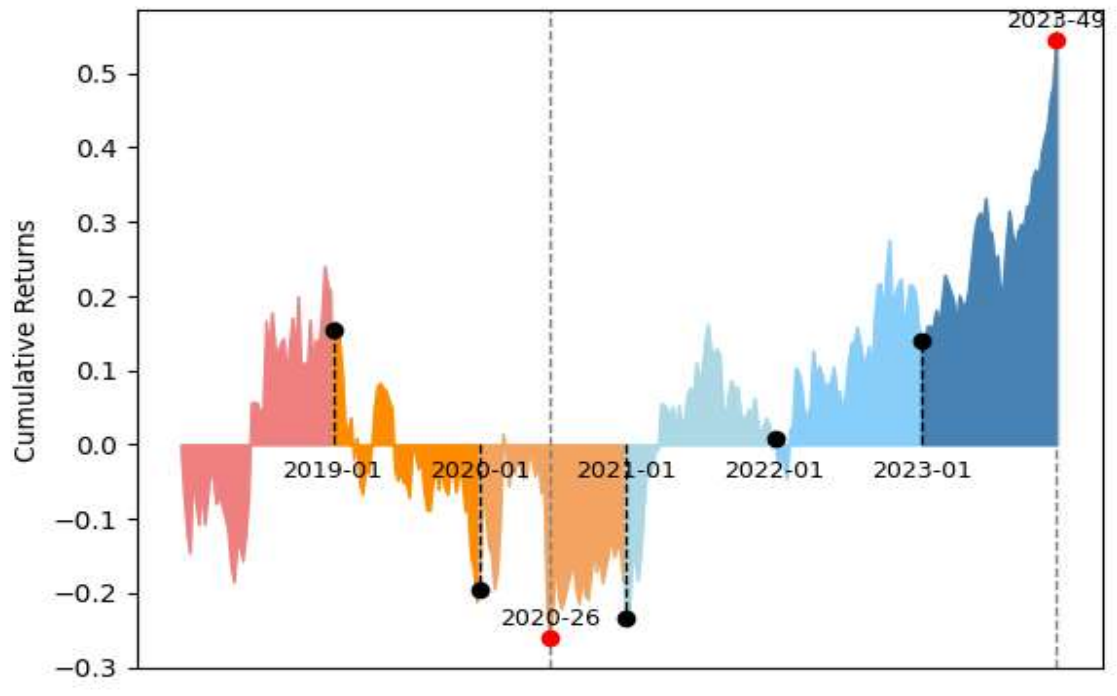

Figure 4 SSE 50 beta-inefficient-momentum

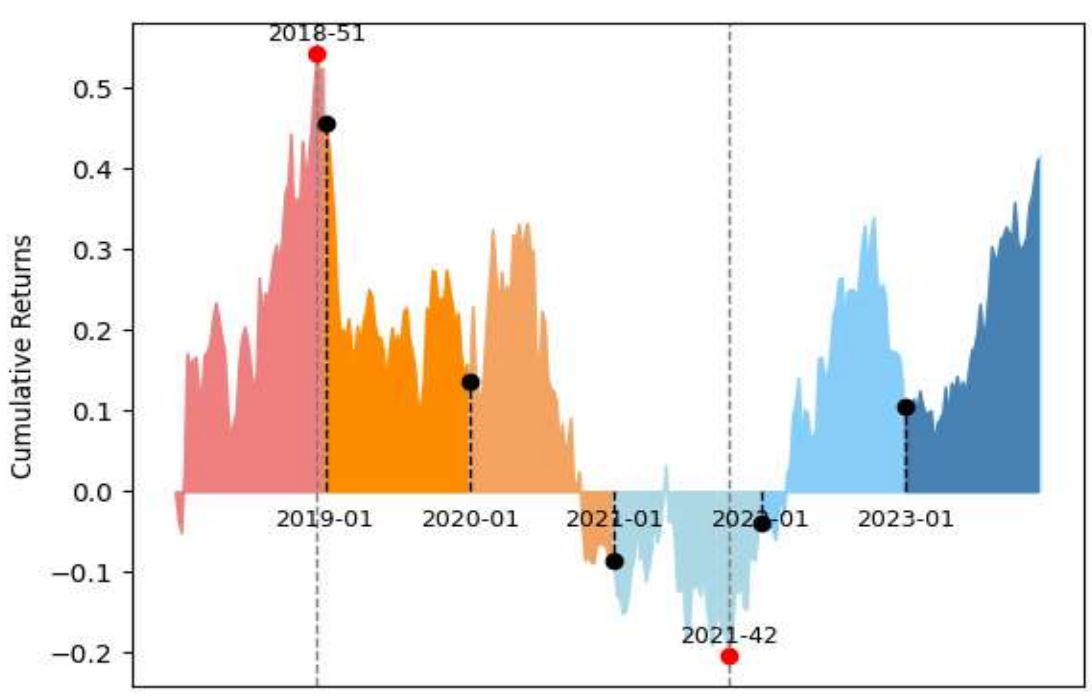

Figure 5 SSE 50 beta-efficient-contrarian

In eq. (10), we can also use the p-index as an input to replace beta. Figure 6 shows that the equally weighted p-index-inefficient portfolio using a momentum strategy performed well, generating annualized rate of return of $9.01\%$. The equally weighted p-index-inefficient portfolio using a contrarian strategy incurred losses, providing annualized rate of return of $-17.2\%$. For equally weighted p-index-efficient portfolios, the contrarian strategy (Figure 7) delivered strong performance ($6.48\%$ annualized rate of return), while the momentum strategy incurred substantial losses ($-9.69\%$ annualized rate of return).

From Figures 2-7, we conclude that among the fifty stocks in the SSE 50 index, efficient (outperforming) stocks failed to sustain their momentum, while inefficient (underperforming) stocks exhibited no mean reversion. Both inefficient-momentum and efficient-contrarian investment strategies yielded significant returns. Among all momentum and contrarian strategies, p-ratio-efficient-contrarian strategy generated the highest annualized rate of return, at $9.97\%$.

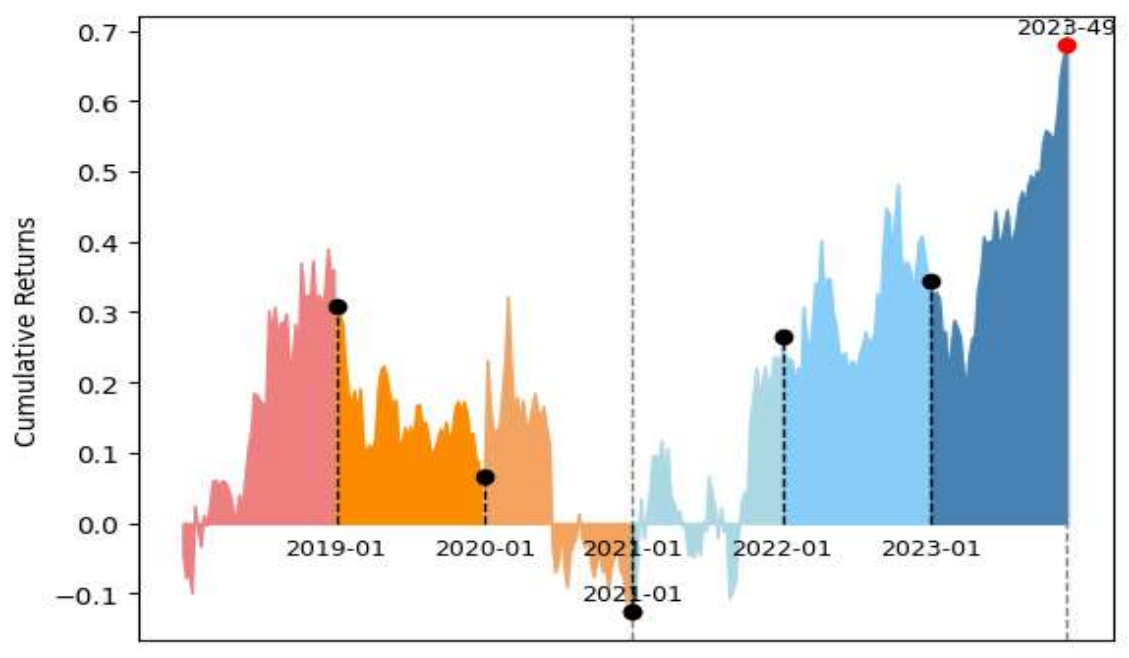

Figure 6 SSE 50 p-index-inefficient-momentum

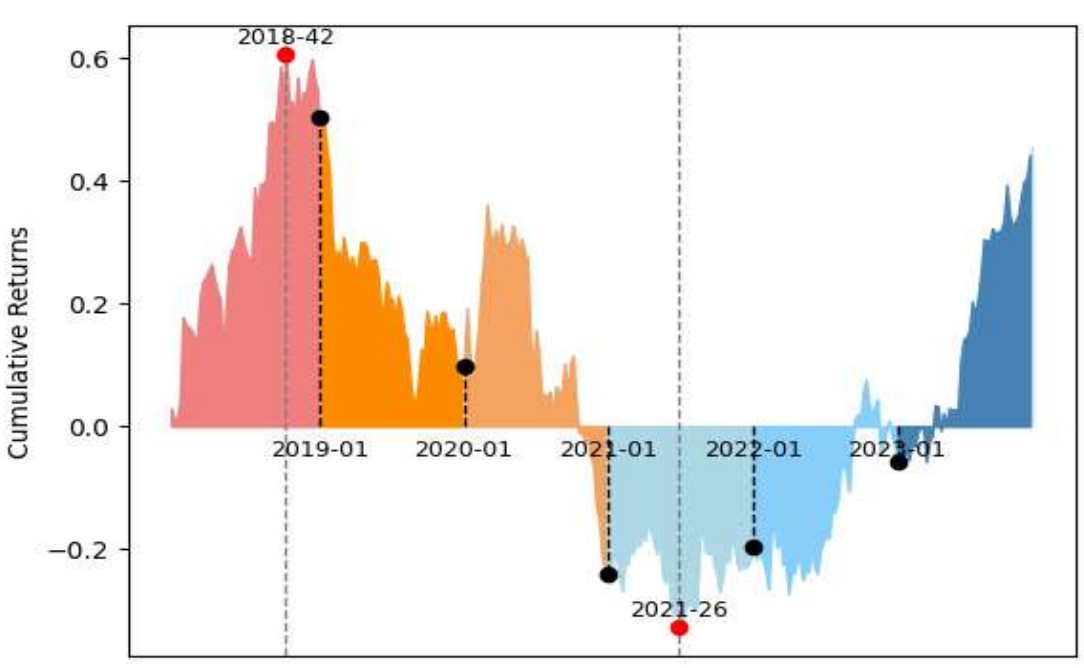

Figure 7 SSE 50 p-index-efficient-contrarian

As shown in Figure 8, the MCIRS method employing the p-index consistently delivered higher returns than its beta-based approach. This may indicate that the p-index is a better risk measure than beta. Figure 9 reveals that the p-ratio-efficiency-contrarian strategy notably surpassed the p-index-inefficiency-momentum and p-index-efficiency-contrarian strategies, emerging as the most robust performer across almost all periods.

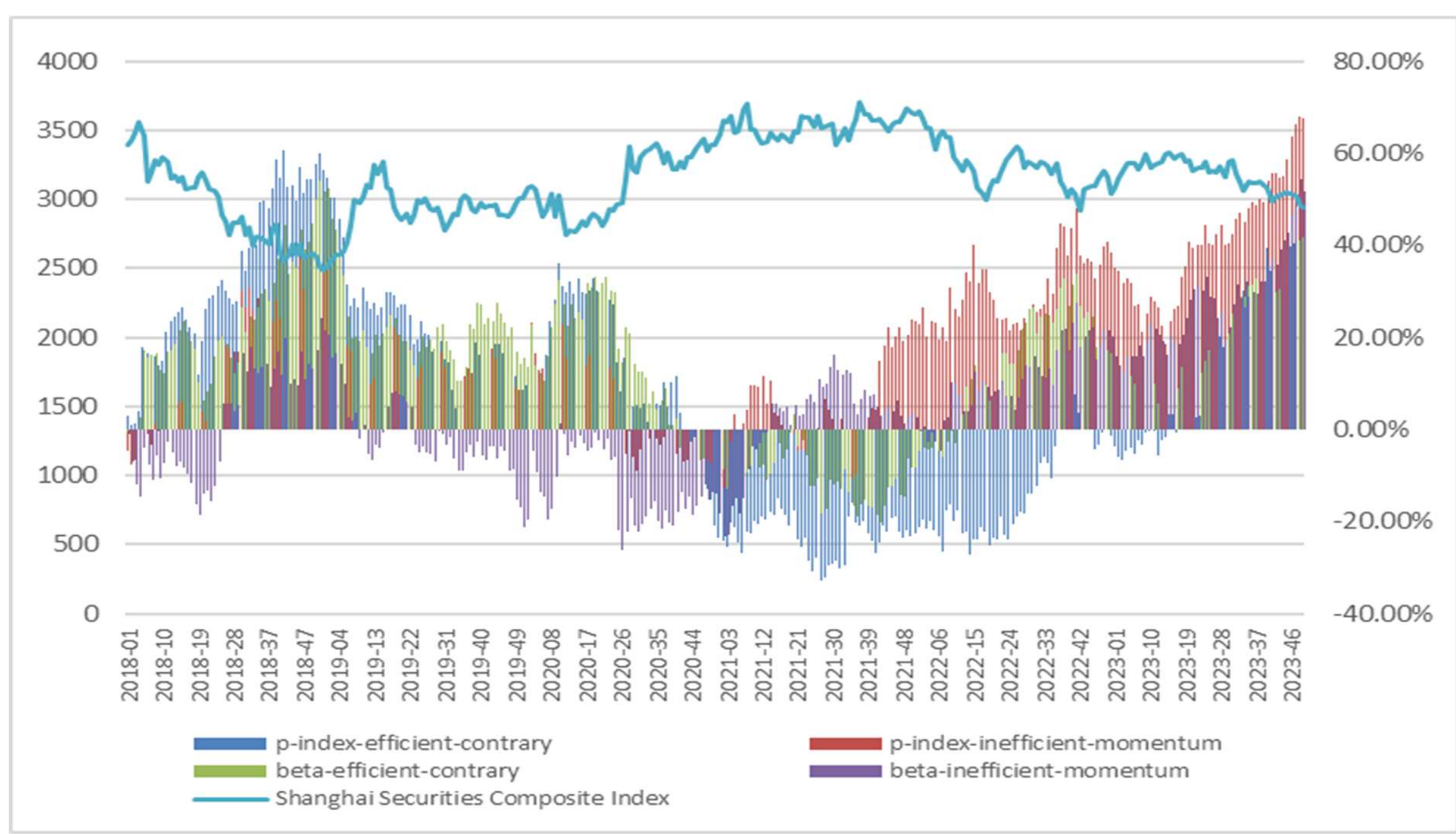

Figure 8 Comparisons of the p-index and beta strategies

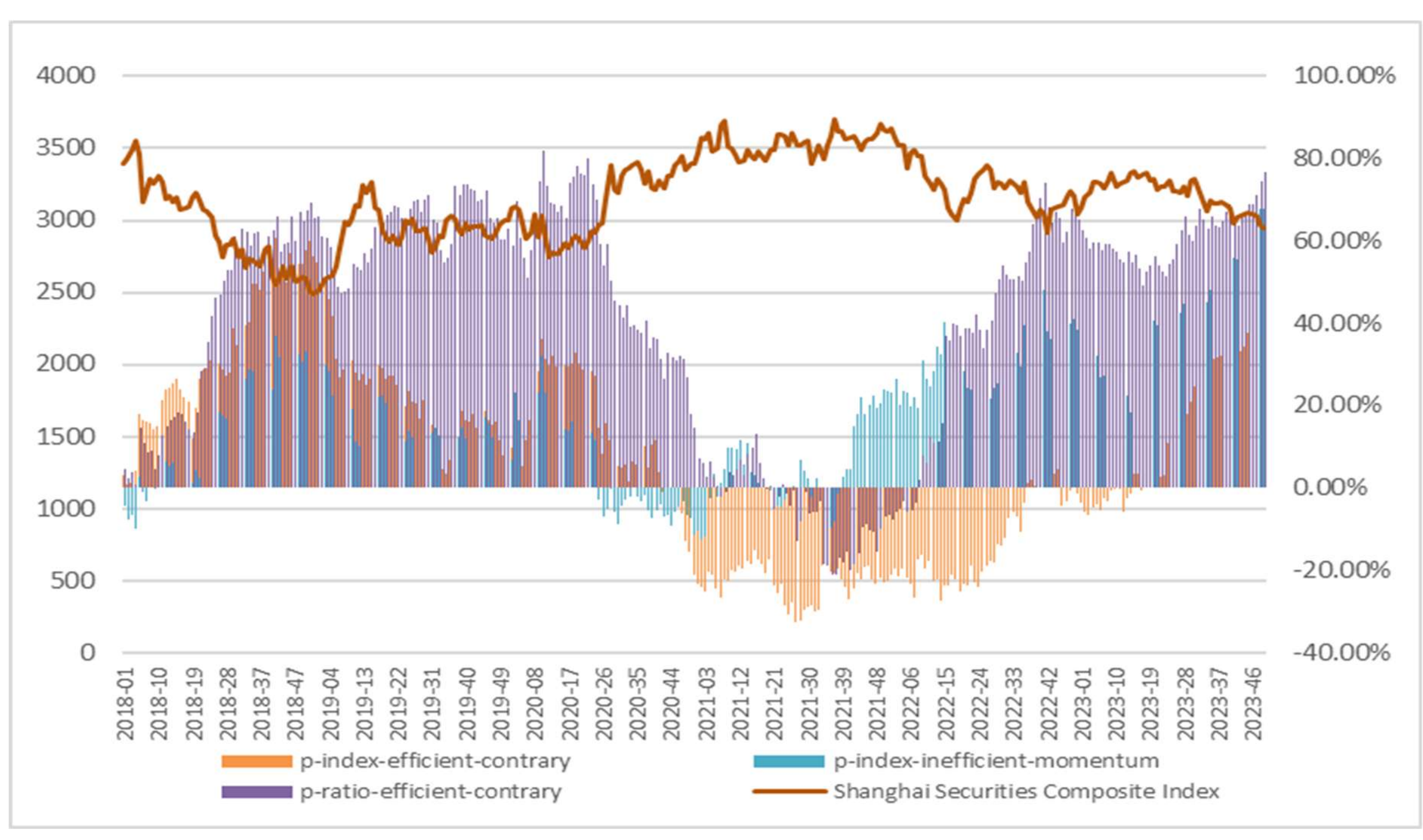

Figure 9 Comparisons of p-ratio-efficient-contrarian, p-index-inefficient-momentum and p-index-efficient-contrarian strategies

Market trading volume can serve as a proxy for investor sentiment, where elevated levels signal heightened market activity and strong sentiment. As demonstrated in Figure 10's synthesis of weekly trading volume data with Figures 6-7, a distinct pattern emerges: The p-index-efficient-contrarian strategy consistently outperformed its p-index-inefficient-momentum counterpart during the 2018-Q2 2020 period (weekly volume: RMB13.3429B), but underperformed thereafter (post-Q2 2020 weekly volume: RMB18.9739B), reflecting a 42% surge in market participation. This inverse relationship reveals that (1) the p-index-efficient-contrarian strategy outperformed in low-sentiment (low-volume) regimes, while (2) the p-index-inefficient-momentum strategy outperformed during high-sentiment (high-volume) periods.

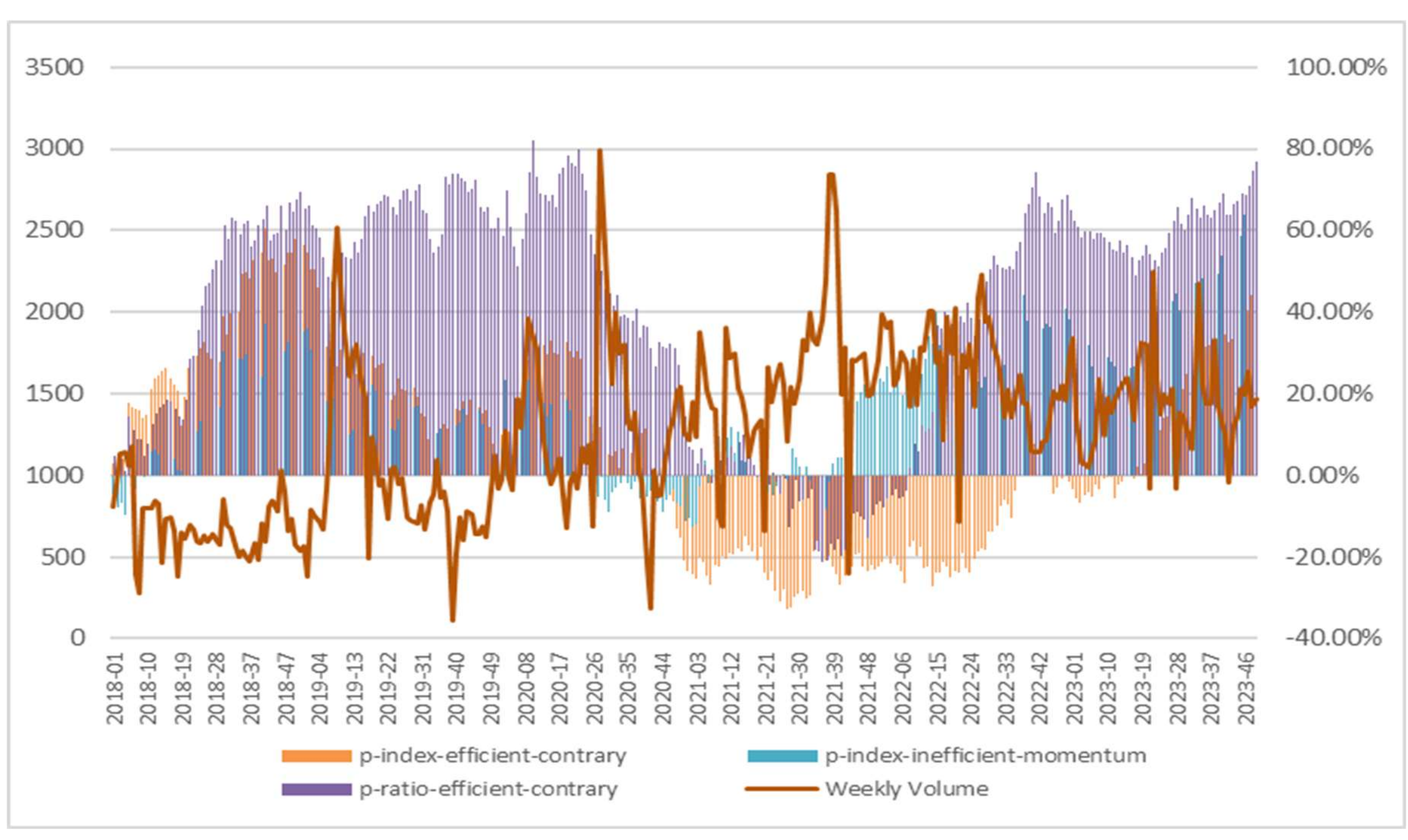

Figure 10 Market trading volume, and p-index-efficient-contrarian vs. p-index-inefficient-momentum strategies

For comparison, we also selected the five hundred stocks comprising the U.S. S&P 500 index from Yahoo Finance for the period from 2018 to 2023. Classify the top 30% of firms by p-ratio (and by p-index and beta in the MCIRS method) as efficient firms and the bottom 30% as inefficient firms. To calculate the p-index and $\pi$ from eq. (8) we use the Dow Jones Total Stock Market Index. As shown in Figures 11–16, efficient stocks (outperforming) maintained momentum, whereas inefficient stocks (underperforming) exhibited mean reversion—contrary to the behavior observed in China's SSE 50 index. The p-index-efficient-momentum strategy produced the highest annualized rate of return (3.69%), followed by the p-ratio-inefficient-contrarian strategy (3.67%) and the beta-efficient-momentum strategy (3.48%).

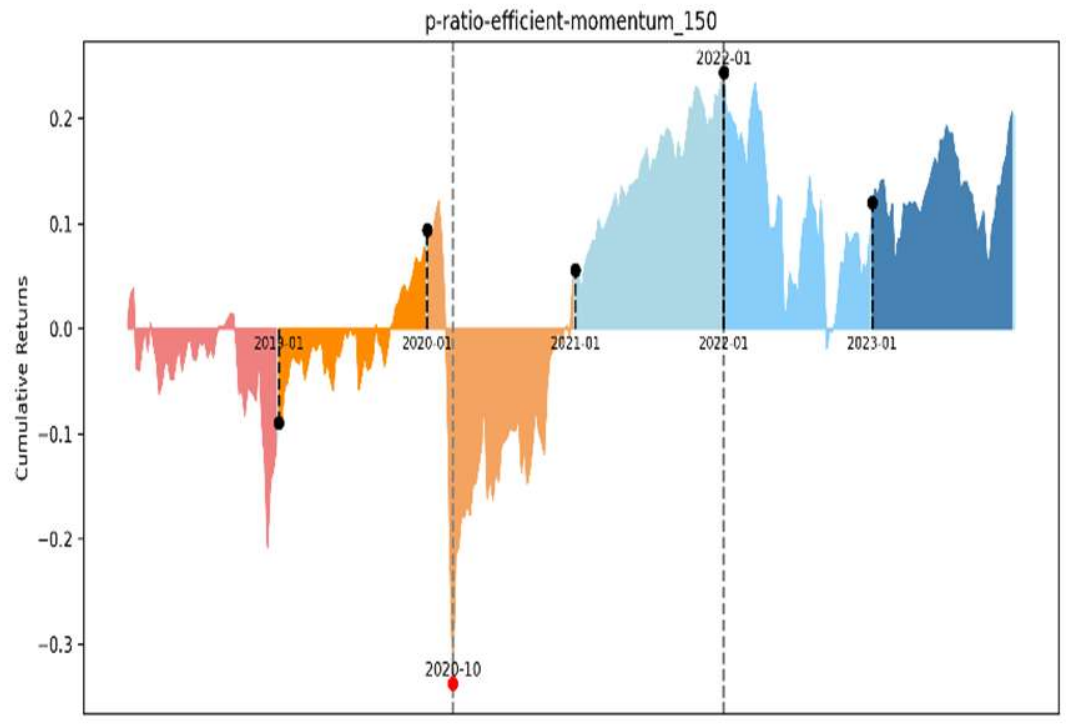

Figure 11 S&P 500 p-ratio-efficient-momentum

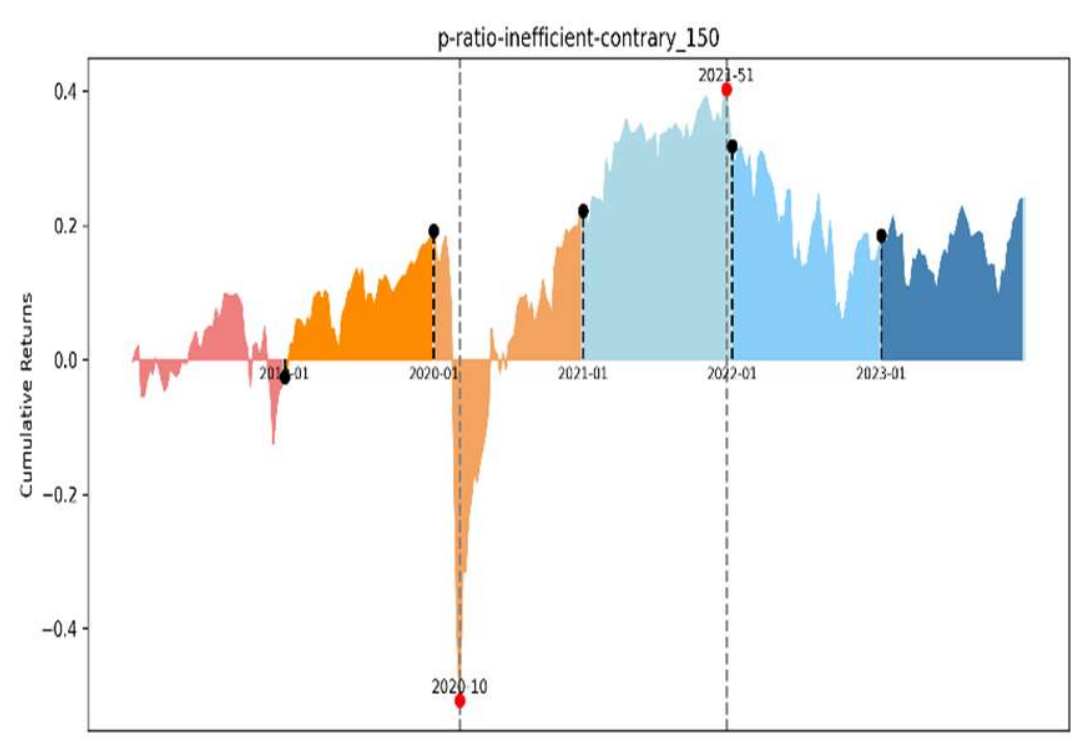

Figure 12 S&P 500 p-ratio-inefficient-contrarian

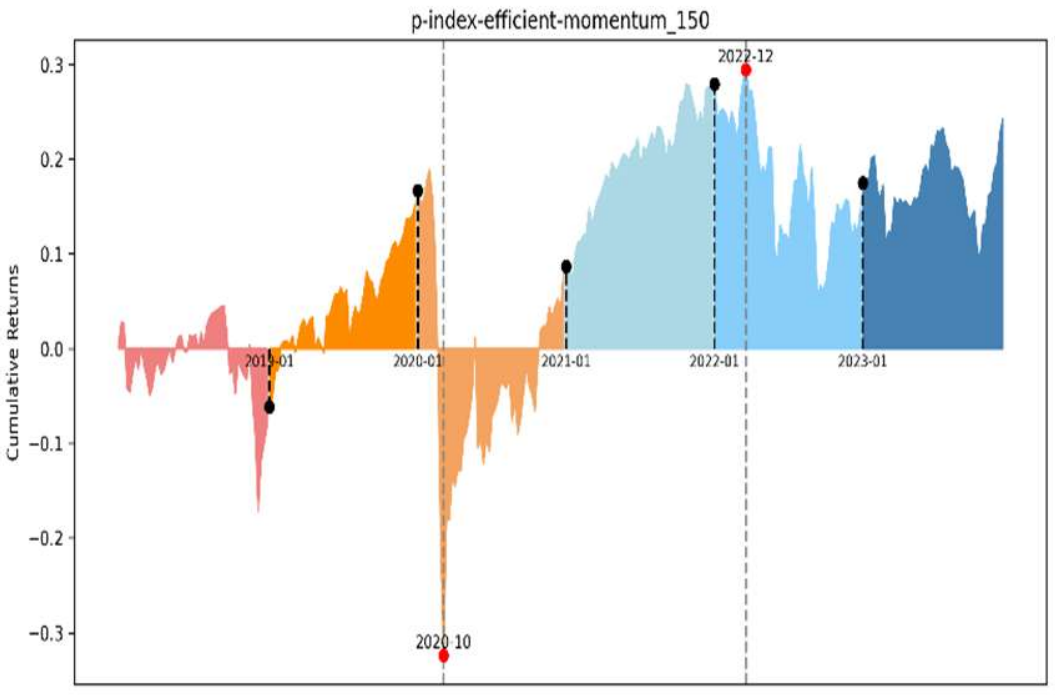

Figure 13 S&P 500 p-index-efficient-momentum

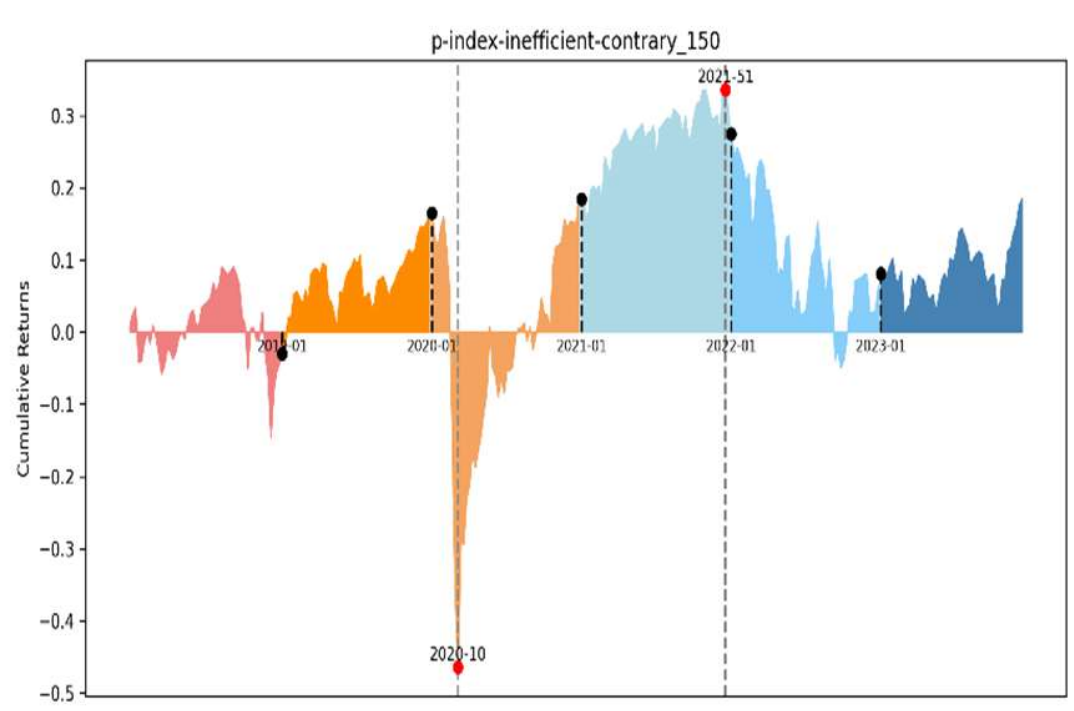

Figure 14 S&P 500 p-index-inefficient-contrarian

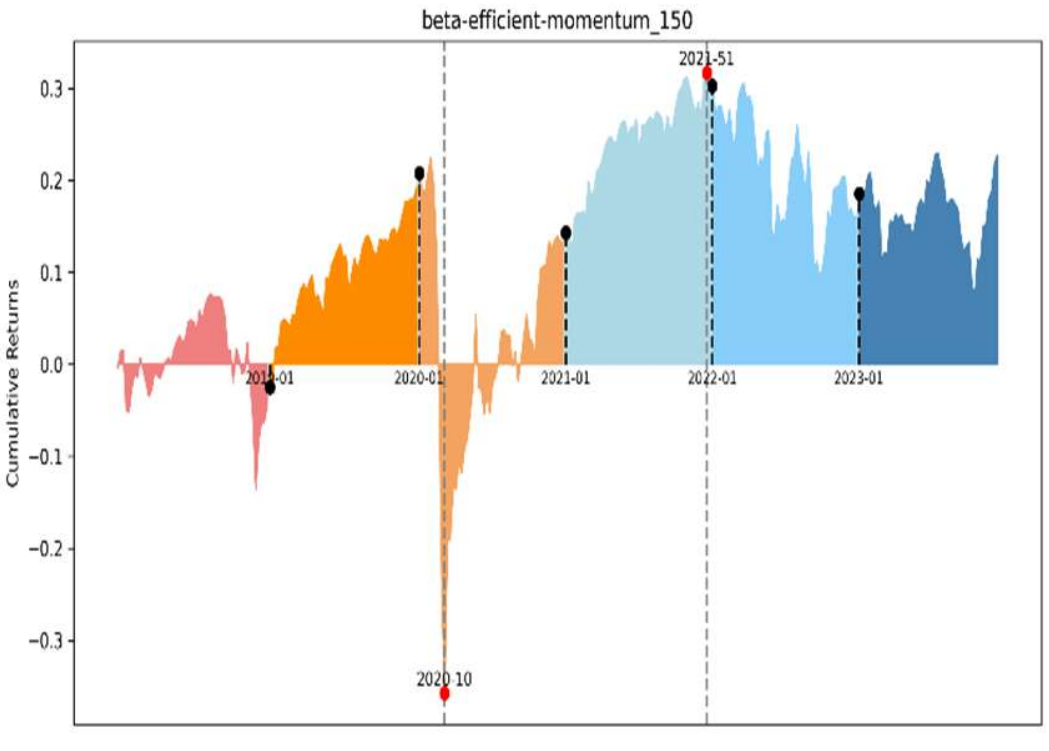

Figure 15 S&P 500 beta-efficient-momentum

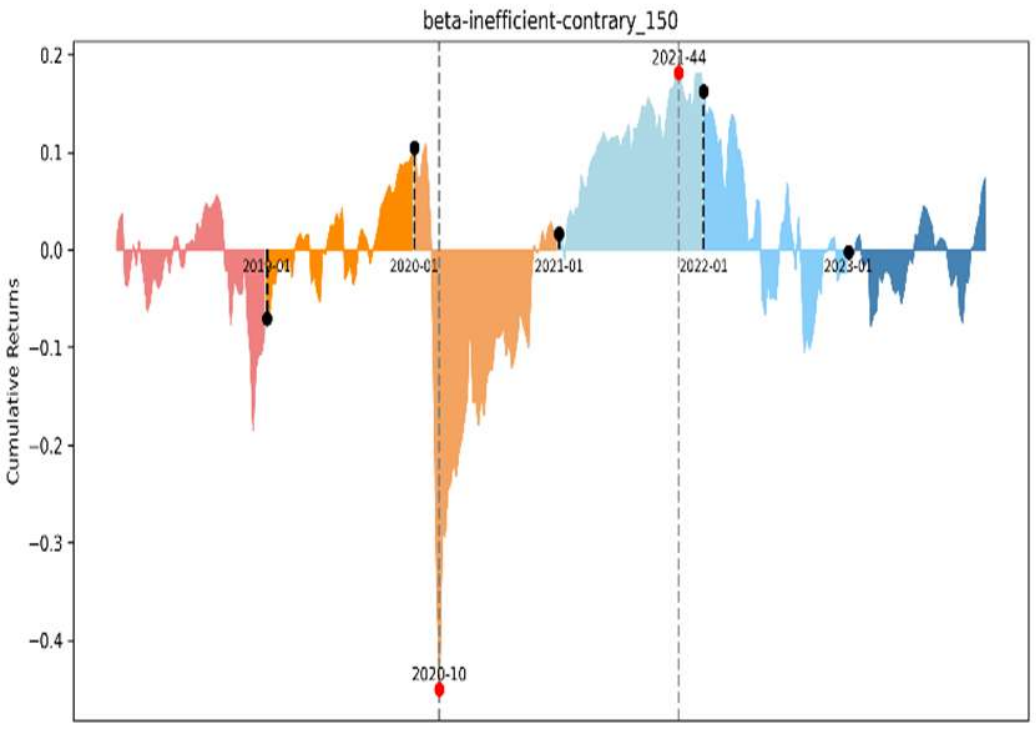

Figure 16 S&P 500 beta-inefficient-contrarian

## 4. Concluding Remarks

The p-index measures the insurance fee for each insured dollar so that the asset $S_0$ can deliver at least a $\delta$ rate of return at $t = T$. Higher p-index means higher risk (i.e., less likelihood) for the asset $S_0$ to deliver at least a $\delta$ rate of return at $t = T$. By using the p-index, we can also derive the p-ratio, a risk-adjusted rate of return. This study sets $\delta$ equal to the risk-free interest rate $(r)$ and evaluates the performance of various investment strategies in the fifty stocks of China's SSE 50 index and the five hundred stocks of the U.S. S&P 500 index from 2018 to 2023.

The empirical results of the fifty stocks of China's SSE 50 index have shown:

First, with the fair price strategy: (a) one-week and one-month holding periods can earn more; (b) the top three stocks with the highest weightings generated annualized rate of returns: 2.31% (one-week period) and 9.78% (one-month period); (c) among seven economic sectors, materials sector stocks generated highest annualized rates of return: 11.04% (one-week period), 11.93% (two-week) and 10.18% (one-month period); and (d) among the fifty stocks in the SSE 50 index, those with high annualized returns exhibited a lower p-index than those with low annualized returns.

Second, for one-week holding period, with momentum and contrarian strategies: (a) the p-ratio-efficient-contrarian strategy produced the highest annualized rate of return (9.97%), followed by the p-index-inefficient-momentum strategy (9.01%) and the p-index-efficient-contrarian strategy (6.48%); (b) the MCIRS method employing the p-index consistently delivered higher returns than its beta-based approach; (c) efficient (outperforming) stocks failed to sustain their momentum, while inefficient (underperforming) stocks exhibited no mean reversion; and (d) by using the market trading volume as a proxy for investor sentiment, it is found that the p-index-efficient-contrarian strategy outperformed in low-sentiment (low-volume) regimes, while the p-index-inefficient-momentum strategy outperformed during high-sentiment (high-volume) periods.

The empirical results of using the five hundred stocks of the U.S. S&P 500 index have shown that efficient stocks (outperforming) maintained momentum, whereas inefficient stocks (underperforming) exhibited mean reversion—contrary to the behavior observed in China's SSE 50 index. The p-index-efficient-momentum strategy produced the highest annualized rate of return (3.69%), followed by the p-ratio-inefficient-contrarian strategy (3.67%) and the beta-efficient-momentum strategy (3.48%).

## Declarations

All authors certify that they have no affiliations with or involvement in any organization or entity with any financial interest or non-financial interest in the subject matter or materials discussed in this manuscript.